\documentclass[manuscript]{aastex}
\begin{document}
\title{Understanding the giant gamma-ray outburst on June 16, 2015 from the blazar 3C 279}
\author{K. K. Singh}
\affil{Physics Department, University of the Free State, Bloemfontein, South Africa 9301 and 
      {Astrophysical Sciences Division, Bhabha Atomic Research Centre, Mumbai, India 400085}}
\author{P. J. Meintjes, F. A. Ramamonjisoa}
\affil{Physics Department, University of the Free State, Bloemfontein, South Africa 9301}
\email{kksastro@barc.gov.in}
\begin{abstract}
A very bright and fast varying gamma-ray flare has been detected from the blazar 3C 279 on June 16, 2015. 
We have studied the broadband spectral energy distribution of the source during the flaring episode and 
in the low activity state using a simple one zone leptonic model. We find that an electron energy distribution 
described by a broken power law can be used to reproduce the broadband emissions during the high and low activity 
states. The flux measurements at radio, infrared and optical frequencies are reproduced by the synchrotron emission 
resulting from the relativistic electrons in a jet magnetic field strength of 0.37 G. The gamma-ray emission from the 
blazar 3C 279 is attributed to the Comptonization of the IR seed photons from the dusty torus with a temperature of 870 K. 
The outburst from the source observed on June 16, 2015 can be ascribed to an efficient acceleration process associated with 
a sudden enhancement in the electron energy density in the emitting region with respect to the low activity state. The 
fast gamma-ray variability at a minute timescale implies that the emission during the flare originates from a more compact 
region and the size of the emission zone in the low activity state is found to be four times larger than that during the flare. 
We have also used the model parameters derived from the broadband spectral energy distribution modelling to investigate a 
few physical properties of the jet during the outburst.
\end{abstract}



\section{Introduction}
The unification scheme of radio-loud active galactic nuclei suggests that blazars exhibit a pair of relativistic 
jets oriented in the opposite directions perpendicular to the accretion disc and one of them is inclined at small 
viewing angles ($\le$ 10$^\circ$) to our line of sight \citep{Urry1995,Padovani2017}. They are powered by accretion 
of matter from the surrounding gas onto supermassive black holes (SMBH) at the centre of their elliptical 
host galaxies. The relativistic jets transport matter in the form of a highly collimated plasma expelled from the nucleus 
of the host galaxies and are extended up to Mpc scales into the intergalactic space. The launching and evolution of 
these powerful relativistic plasma jets are not completely understood, however, it is generally believed that they are 
powered either by the rotational energy of a spinning SMBH at the centre through the Blandford-Znajek process or 
by the gravitational energy of the accreting matter \citep{Blandford1977}. The energy flow is dissipated through the 
multi-wavelength emission downstream along the jet. The broadband non-thermal emissions from the plasma 
jets are amplified due to the relativistic Doppler boosting. The low energy emission observed at radio-optical and 
X-ray energies is attributed to the synchrotron radiation due to the relativistic electrons in the jet magnetic field. 
The high energy emission from blazars is still under debate and different scenarios have been proposed to explain 
the observed $\gamma$-ray spectral features \citep{Bottcher2013}. 
\par
In the leptonic processes, $\gamma$-ray photons are produced by the inverse Compton (IC) scattering of low energy seed 
photons by the relativistic electrons in the blazar jet \citep{Dermer1995,Dermer2009}. The energy of the target photon 
determines the dominance of Comptonization in the Thomson or Klein-Nishina (K-N) regime. If $\gamma$-rays are produced by 
the IC scattering in the Thomson regime, their spectral shape is straightforward related to the energy distribution of 
relativistic electrons in the emitting region. Whereas, the relation between the observed $\gamma$-ray spectrum and energy 
distribution of electrons is very complex if the $\gamma$-ray production takes place in the K-N regime. Therefore, the 
spectral shapes of the observed high energy $\gamma$-rays from blazars are an indirect probe of the particle acceleration 
mechanisms in the jet, which have not been properly understood. The leptonic models with different forms of electron 
energy distribution cooling down in various populations of low energy target photons (synchrotron photons produced  
in the emission region, accretion disc photons reprocessed by the broadline region and dusty torus external to the jet) 
are widely used to model the broadband spectral energy distribution (SED) of most blazars 
\citep{Bottcher2013,Abdo2010a,Agudo2011,Singh2018}. Hadronic processes including proton synchrotron, muon synchrotron, 
pion decay and cascade emission are also invoked to model the $\gamma$-ray emission from some 
blazars \citep{Aharonian2002,Bottcher2009,Cerruti2015}. Recent detection of astrophysical neutrino events from the 
direction of the blazar TXS 0506+056 provides additional evidence for the hadronic processes involved in the 
$\gamma$-ray production \citep{Aartsen2018a,Aartsen2018b}. The $p-p$ interactions are generally not considered in the 
hadronic processes for $\gamma$-ray emission from blazars due to a low particle density in the emitting zone. In a quite 
different scenario, the high energy $\gamma$-rays are not produced in the blazar jet but in the intergalactic space by 
the protons (components of ultra high energy cosmic rays accelerated in the jet) escaping from the emission region and 
beamed in the direction of the observer. These energetic protons initiate an electromagnetic cascade in the intergalactic 
space via interaction with low energy photons from the extragalactic background light (EBL) or cosmic microwave background 
radiation (CMBR) through photo-meson and pair-creation processes \citep{Essey2011,Murase2012,Takami2013}. The high energy 
$\gamma$-ray photons are produced by the electromagnetic cascade and propagate towards the Earth without any significant attenuation. 
This is known as hadron beam scenario and has been used to explain the $\gamma$-ray emission from a sample of hard spectra blazars 
having extreme properties \citep{Tavecchio2019,Singh2019a}.
\par
Multi-wavelength emissions from the blazars are observed to exhibit a rapid variability on minute timescales during the flaring episodes 
associated with a dramatic increase in the luminosity \citep{Albert2007,Singh2012,Singh2015,Bartoli2016,Singh2019b} and a significant 
drop in the degree of optical polarization \citep{Abdo2010b,Singh2019c}. The behaviour of the spectral and temporal changes during  
near simultaneous multi-wavelength or orphan flares is found to be erratic and has yet not been understood. Various time-dependent 
single-zone emission models are used to explain the observed properties of outbursts from the blazars \citep{Chen2011,Mastichiadis2013,
Singh2017}. Multi-zone emission models have also been proposed to explain the stochastic nature of the variability during the 
flaring activity of blazars \citep{Zhang2016}. Both single-zone and multi-zone models involving leptonic and hadronic processes 
are extensively used to study the flares depending on the observed properties of a source. However, a clear understanding of the 
physical process is lacking. Therefore, exploring the physical mechanism involved in the non-thermal emission during the flaring 
activity of blazars remains an important and major task in the blazar research today. In this work, we focus on the historical 
$\gamma$-ray flare observed from the blazar 3C 279 on June 16, 2015. 3C 279 is one of the most distant blazars at a 
cosmological redshift $z$= 0.536 and belongs to the flat spectrum radio quasar (FSRQ) subclass of the blazars \citep{Burbidge1965}. 
The FSRQ type blazars are characterized by the strong broad emission lines in the optical spectrum and a bright accretion disc due 
to high accretion rate. The high energy radiation from FSRQs is generally produced by IC scattering of the external seed photons 
emitted from the bright accretion disc and reprocessed by the broadline region and extended dusty torus surrounding the central SMBH.
In the present work, we use IR photons from the dusty torus to reproduce the observed $\gamma$-ray emission from the FSRQ blazar 3C 279 
during the giant outburst on June 16, 2015. In Section 2, we briefly summarize the important results from the June 2015 flare of 3C 279.
The multi-wavelength data set and daily light curve are described in Section 3 and Section 4 respectively. In Section 5, we 
explain the single zone leptonic model to reproduce the broadband SEDs of the source. Finally, we discuss the important findings 
in Section 6 and conclude the study in Section 7. 
We have adopted the cosmological parameters H$_0$= 70 km s$^{-1}$ Mpc$^{-1}$, $\Omega_m$= 0.27 and $\Omega_\Lambda$= 0.73 from 
the $\Lambda$CDM model.

\section{Summary of June 2015 outburst from 3C 279}
A giant outburst from the blazar 3C 279 was detected by the \emph{Fermi}-Large Area Telescope (LAT) with the peak flux level 
of (3.6$\pm$0.2)$\times$ 10$^{-5}$ ph cm$^{-2}$ s$^{-1}$ above 100 MeV on June 16, 2015 \citep{Ackermann2016}. It was found  
historically to be the highest $\gamma$-ray flux measured from 3C 279 with an isotropic luminosity close to 10$^{49}$ erg s$^{-1}$. 
This active phase of the source allowed the investigation of $\gamma$-ray variability on minute-scale with a flux doubling time 
of $\sim$ 5 minutes. This suggests a very compact emission region in the jet of 3C 279. The high energy $\gamma$-ray spectrum 
measured by LAT during the flaring activity of the source was more favorably described by a log-parabola model with   
spectral and curvature indices of 1.71$\pm$0.09 and 0.20$\pm$0.05 respectively than the power law model having a spectral index 
of 2.01$\pm$0.05. The highest energy of photons detected was 56 GeV with a probability of 99.99$\%$ at the end of the flare 
from 3C 279. During the flare, the high energy component of the SED peaks at energies between 0.3 GeV and 1 GeV, whereas 
the peak energy is 1 GeV at the beginning and end of the flare due to relatively hard spectra. This is not consistent with the 
previous emission states of 3C 279 when the peak energy in SED has been observed below 0.1 GeV. 
The instruments on board the AGILE satellite also detected intense $\gamma$-ray emission in the energy range 30 MeV-50 GeV from 
3C 279 during the outburst \citep{Pittori2018}. The high energy $\gamma$-ray flux measured by AGILE above 100 MeV was 
(1.3$\pm$0.13)$\times$ 10$^{-5}$ ph cm$^{-2}$ s$^{-1}$ with a power law photon spectral index of 2.14$\pm$0.11 during the flare. 
The results obtained from AGILE observations during the high activity state of the blazar 3C 279 are compatible with 
the \emph{Fermi}-LAT values within statistical uncertainty. The high energy $\gamma$-ray flare was also followed up with the 
ground based H.E.S.S. telescopes and resulted in a signal with 8.7$\sigma$ statistical significance above an energy threshold 
of 66 GeV in $\sim$ 2.2 hours of observation \citep{Hess2019}. 
The photon spectrum observed by  H.E.S.S. telescopes was described by a soft power law with a spectral index of 4.2$\pm$0.3 and 
the average integral flux above 200 GeV was found to be (7.6$\pm$0.7)$\times$ 10$^{-12}$ ph cm$^{-2}$ s$^{-1}$. No indication of 
statistically significant variability was detected in the light curve and the flux points above 100 GeV were consistent with an 
average flux level of (6.5$\pm$0.6)$\times$ 10$^{-11}$ ph cm$^{-2}$ s$^{-1}$ during the night of outburst. 
Contemporaneous X-ray and optical/UV observations also hint for the flaring activity of 3C 279 at lower 
energies \citep{Pittori2018,Hess2019}. The X-ray flux measured by the X-ray Telescope (XRT) on board the 
\emph{Neil Gehrels Swift} observatory in the energy range 0.3-10 keV during the peak was about 4 times the flux level in 
low activity state of the source with a correlated variability in high energy $\gamma$-rays. The optical emission in R band was 
also observed to increase by 40$\%$ during the outburst \citep{Hess2019}.

\section{Data Set}
As discussed in Section 2, the high energy $\gamma$-ray emission from the blazar 3C 279 started to increase from June 14, 2015, 
attained a peak on June 16, 2015 and finally decreased to a low activity state on June 17, 2015. Therefore, we have used near 
simultaneous broadband data available from various space and ground based observations of the blazar 3C 279 during June 1-30, 2015 
for the SED modelling. We have divided the whole period into three epochs: Pre-flare (June 1-13, 2015), Flare (June 16, 2015) 
and Post-flare (June 18-30, 2015). We have also obtained the multi-wavelength light curves from high energy $\gamma$-ray to radio 
between May 1, 2015 and July 31, 2015 (MJD 57143-57234) to study the emission behaviour of the source in different energy bands 
for a longer period. The details of the multi-wavelength dataset used in this study are described below.  

\subsection{$\gamma$-ray}
We have analyzed high energy $\gamma$-ray data from the \emph{Fermi}-LAT observations in the energy range 100 MeV - 500 GeV using 
standard Fermitools (Fermi 1.0.1). The LAT is a pair-conversion telescope which directly detects $\gamma$-rays in the energy 
range from 20 MeV to more than 1 TeV \citep{Atwood2009,Ajello2017}. The publicly available Pass8 event data on blazar 3C 279 was 
downloaded from the \emph{Fermi}-LAT data server\footnote{https://fermi.gsfc.nasa.gov/cgi-bin/ssc/LAT/LATDataQuery.cgi}. 
We have only selected source class events with evclass = 128 and evtype = 3 in a region of interest (ROI) of radius 10$^\circ$ 
centered at the position of 3C 279 (4FGL J1256.1-0547). All point sources listed in the 4FGL catalog (gll$\_$psc$\_$v18.fits) and 
located within 20$^\circ$ radius from the center of ROI along with the Galactic (gll$\_$iem$\_$v06.fits) and extragalactic isotropic 
diffuse emission (iso$\_$P8R3$\_$SOURCE$\_$V2.txt) for the background subtraction are included in the model file for an unbinned 
likelihood analysis. We have produced the time averaged spectral flux points in the energy range 100 MeV - 500 GeV for the three 
epochs of 3C 279 observations using $gtlike$. The source 3C 279 is modelled with a log-parabola spectrum whereas the spectra of 
all the point sources in the ROI have been kept the same as in the 4FGL catalog \citep{4FGL2019}.

\subsection{X-ray}
The archival X-ray data available during three epochs in the energy bands 15-50 keV and 2-20 keV have been obtained from the 
Burst Alert Telescope (BAT) and Monitor of All-sky X-ray Image (MAXI) observations of the blazar 3C 279 respectively. The BAT 
transient monitor \citep{Krimm2013} on board the \emph{Neil Gehrels Swift} observatory provides online data from the 
daily observations of the transient events in the sky\footnote{https://swift.gsfc.nasa.gov/results/transients}. We have  
estimated the time averaged flux points in the energy range 15-50 keV for three epochs of 3C 279 observations. The instruments 
on the MAXI experiment also detect the X-ray transient events and monitor the intensity fluctuations of the known sources 
in the energy range 2-20 keV \citep{Matsuoka2009}. We have used MAXI on-demand process\footnote{http://maxi.riken.jp/mxondem} 
to get the data for the blazar 3C 279 in two energy bands 2 keV-6 keV and 6 keV-20 keV. Thus data from MAXI and BAT observations 
provide three spectral points for the X-ray component of the broadband SED of the source.   

\subsection{Optical and Radio}
The optical/IR contemporaneous data during three epochs have been obtained from the Small \& Moderate Aperture Research Telescope 
System (SMARTS) project \footnote{http://www.astro.yale.edu/smarts}. The SMARTS project provides daily observations of selected 
blazars using two small aperture telescopes in three optical (B, V \& R) and two infrared (J \& K) bands. The details of data reduction 
and analysis are described in \citep{Bonning2012}. We have used the publicly available data in magnitude scale to get the energy flux 
points in different bands following the conversion described in \citep{Hayashida2015}. The radio data at 15 GHz are obtained from 
the 40 m telescope at Owens Valley Radio Observatory (OVRO)\footnote{http://www.astro.caltech.edu/ovroblazars}. The details of  
the radio monitoring program using the 40 m telescope are described in \citep{Richards2011}. We have used the publicly available data of  
the blazar 3C 279 during the above period.

\section{Multi-wavelength Light Curve}
\begin{figure}
\epsscale{0.75}
\plotone{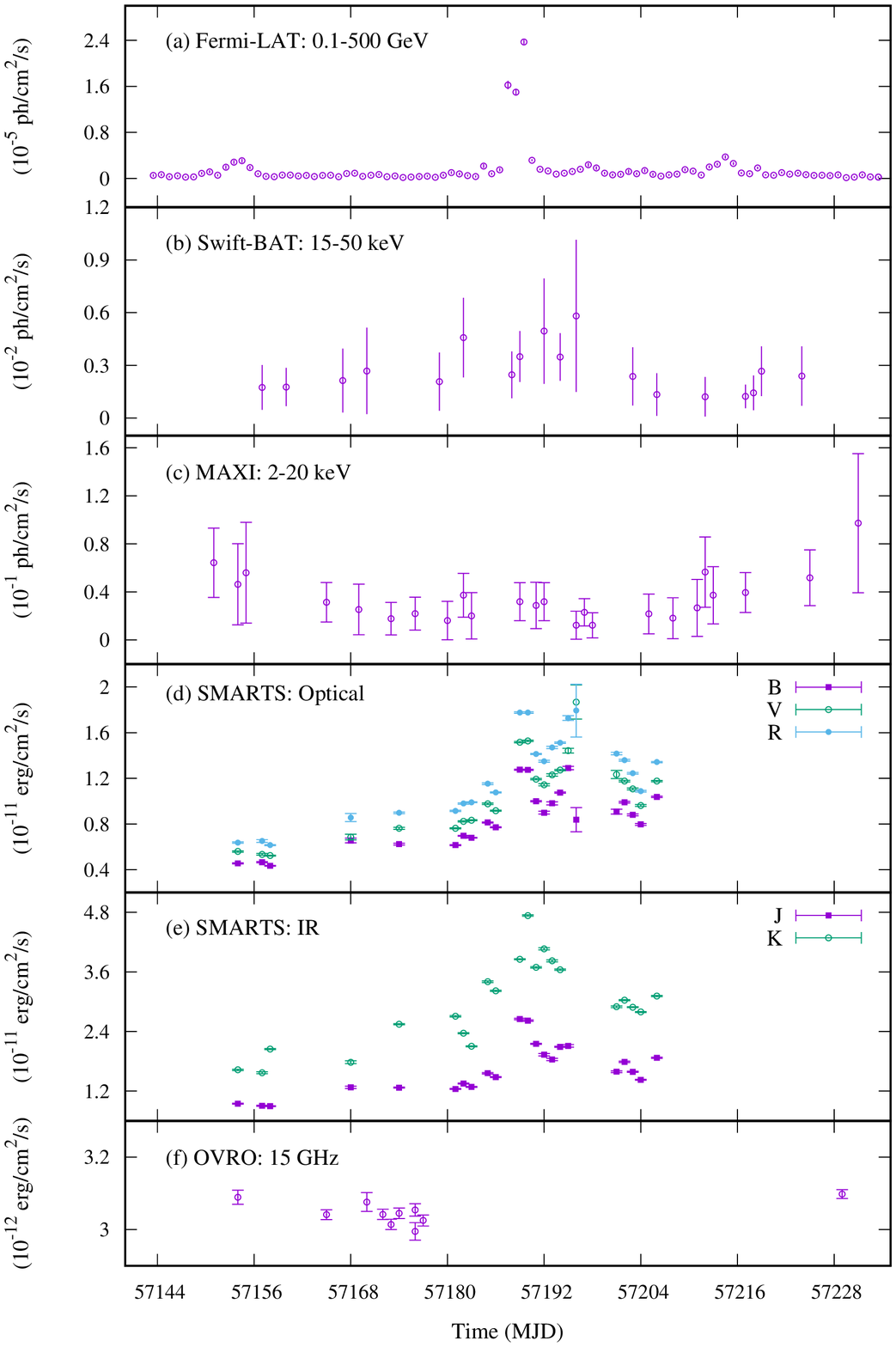}
\caption{One day binned multi-wavelength light curve of the FSRQ 3C 279 during the period May 1, 2015 
	to July 31, 2015 (MJD 57143-57234) from the \emph{Fermi}-LAT, \emph{Swift}-BAT, MAXI, SMARTS 
	and OVRO-40 m telescope observations. The giant $\gamma$-ray flaring episode peaks on June 16, 2015 
	(MJD 57189).} 
\label{fig1}
\end{figure}

The multi-wavelength light curve of the blazar 3C 279 during May 1, 2015 - July 31, 2015 (MJD 57143-57234) is shown in 
Figure \ref{fig1}(a-f). The high energy $\gamma$-ray flux points in Figure \ref{fig1}(a) correspond to the one day 
binning of the \emph{Fermi}-LAT data in the energy range 0.1-500 GeV. All the flux values have been derived at 
a statistical significance of 3$\sigma$ and above assuming a power law spectrum of the source. The X-ray data points plotted in 
Figure \ref{fig1}(b \& c) are obtained from the daily \emph{Swift}-BAT and MAXI observations in the energy range 15-50 keV and 
2-20 keV respectively. The error bars in both the light curves correspond to the statistical uncertainty  at 1$\sigma$ level. 
Near simultaneous daily flux measurements in the optical/IR (B, V, R, J \& K) bands and at 15 GHz radio from the SMARTS and OVRO 
observations are reported in Figure \ref{fig1}(d-f) respectively. A visual inspection of the one day binned multi-wavelength 
light curve suggests that the flaring activity of the blazar 3C 279 during June 14-16, 2015 (MJD 57187-57189) is more 
prominent  in the high energy $\gamma$-ray band than in the remaining lower energy bands. In fact, the X-ray emissions in 
the 15-50 keV and 2-20 keV energy bands do not show any noticeable change during the $\gamma$-ray outburst. 
The optical/IR light curves indicate an increasing activity in the emission level in all five bands (B, V, R, J \& K) during 
the $\gamma$-ray flaring episode. The radio emission measured at 15 GHz is consistent with the low activity state of the source. 
Therefore, we observe that the giant $\gamma$-ray flaring activity of the FSRQ 3C 279 during June 14-16, 2015 (MJD 57187-57189) 
with a peak on June 16, 2015 (MJD 57189) is accompanied by a simultaneous increase in the optical/IR emission level. 
In order to quantify the degree of multi-band cross correlation, we have estimated the Spearman's rank correlation 
coefficient ($r_s$) using simultaneous flux measurements in two different energy bands. For high energy $\gamma$-ray 
and optical/IR (B, V, R, J \& K) light curves, the estimated values of $r_s$ are in the range 0.60-0.72 at a statistical 
significance level of more than 95$\%$. This implies a strong positive correlation between $\gamma$-ray and optical/IR emissions 
from the blazar 3C 279. The high energy $\gamma$-ray flux points show a weak correlation ($r_s =$0.15-0.21) with X-ray emission 
in 15-50 keV and 2-20 keV bands only 50$\%$ statistical significance level. The correlation between X-ray and optical/IR emissions 
is also found to be weak. The radio emission is observed to be very weakly correlated ($r_s =$0.08) with the high energy $\gamma$-ray 
flux measurements. Therefore, the correlation analysis based on the Spearman's rank correlation suggest that the high energy $\gamma$-ray 
and optical/IR emissions from the FSRQ 3C 279 are physically connected. This also hints support for the leptonic emission model 
for the blazars.


\section{Spectral Energy Distribution Modelling}
We have invoked a simple one zone leptonic model to reproduce the broadband emissions observed from the blazar 3C 279 during 
the period June 1-30, 2015. In this model \citep{Massaro2006,Tramacere2009,Tramacere2011}, the non-thermal emission is assumed 
to originate from a spherical blob of radius $R$ moving down the jet with a bulk Lorentz factor $\Gamma$ and entangled with a uniform 
magnetic field of strength $B$. The relativistic motion of the emitting blob leads to the strong Doppler boosting of the observed 
radiation due to the beaming effect. The relativistic Doppler factor $\delta$ is given by 
\begin{equation}
	\delta~=~\frac{1}{\Gamma~(1-\beta~\rm cos \theta)}
\end{equation}	
where $\theta$ is the viewing angle with respect to the line of sight of the observer and $\beta$ is the bulk speed of the blob in units 
of the speed of light $c$ in vacuum. Under the small viewing angle approximation for blazars ($\theta \sim 5^\circ$ for 3C 279), 
$\delta \approx \Gamma$. The blob is homogeneously filled with the plasma of relativistic leptons ($e^\pm$) and cold protons (expected 
to be present for providing charge neutrality). The protons are assumed to be at rest in the comoving frame and only relativistic leptons 
contribute to the non-thermal emission of the blob. The electrons are accelerated to relativistic energies through putative acceleration 
processes (e.g Fermi first and second order) and then injected to the blob. The electrons in the blob cool via the radiative processes 
namely synchrotron emission and inverse Compton scattering. The differential electron distribution is given by a broken power law, 
which can be expressed as 
\begin{equation}
	\begin{array}{lcl}
		n(\gamma) & = & N K \gamma^{-p}~~,~~\gamma_{min}~\le~\gamma~\le~\gamma_b\\
		          & = & N K \gamma_b^{q-p} \gamma^{-q}~~,~~\gamma_b~\le~\gamma~\le~\gamma_{max}
        \end{array}
\end{equation}
where $N$ and $\gamma~(=\frac{E_e}{m_e c^2})$ are the number density and Lorentz factor of the electrons respectively. 
$p$ and $q$ are the low and high energy spectral indices respectively. $\gamma_{min}$, $\gamma_b$ and $\gamma_{max}$ are 
the Lorentz factors corresponding to the minimum, break and maximum energies of the electron population. The electron distribution 
is normalized through the constant $K$ as 
\begin{equation}
	\int_{\gamma_{min}}^{\gamma_{max}} \frac{n(\gamma)}{N}~d\gamma~=~1
\end{equation}	
The broadband emission from the blazar 3C 279 observed at radio and optical frequencies is dominated by the synchrotron radiation 
of electrons described by Equation (2) interacting with the tangled magnetic field of the blob. Whereas, the high energy emission 
at X-ray and $\gamma$-ray energies is reproduced by Comptonization of the seed photons originating from the dusty torus (dust present 
in the cold molecular gas). The dusty torus (DT) is considered as an isotropic structure obscuring the accretion disc surrounding 
the SMBH. This is expected to survive at locations with ambient temperature below 1500 K surrounding the blazar. The location of 
DT from the central SMBH ($R_{DT}$) can be expressed as \citep{Sikora2002}
\begin{equation}
	R_{DT}~\approx~\frac{1}{T_D^2}\left(\frac{L_{disc}}{4\pi\sigma_{SB}}\right)^{1/2} 
\end{equation}	
where $T_D$, $L_{disc}$ and $\sigma_{SB}$ are the dust temperature, accretion disc luminosity and Stefan-Boltzmann constant respectively. 
The dust is heated by the thermal emission at UV/optical wavelengths from the accretion disc and re-emits at IR wavelengths. The disc 
emission reprocessed by DT (IR photons) acts like an external radiation field for the inverse Compton scattering of the seed photons by 
the relativistic electrons described by Equation (2). The Comptonization of IR seed photons emitted from the DT is assumed to take place 
in the Thomson regime. Assuming that the intensity of the radiation produced by synchrotron ($I_{syn}$) and inverse Compton ($I_{IC}$) 
processes is isotropic, the total observed flux is calculated with \citep{Begelman1984}
\begin{equation}
	F_{obs}(\nu_{obs})~=~\frac{\delta^3 (1+z)V}{d_L^2}\left(I_{syn}(\nu) + I_{IC}(\nu)\right)
\end{equation}	
where $V$ is the volume of the emission region and $d_L$ is the luminosity distance (in $\Lambda$CDM model) of the source at redshift $z$. 
The observed frequency $\nu_{obs}$ of the radiation is related to the emitted frequency $\nu$ as 
\begin{equation}
	\nu_{obs}~=~\frac{\delta \nu}{1+z}
\end{equation}	
This simple one zone leptonic model for the blazar emission uses $R$, $z$, $\Gamma$, $\theta$,
$B$, $N$, $\gamma_{min}$, $\gamma_{max}$, $p$, $q$, $L_{disc}$, $R_{DT}$ and $T_D$ as input parameters. 
These parameters sufficiently  specify the properties of non-thermal broadband emission region 
and electron energy distribution in the jet. Therefore, the observed multi-wavelength data can be 
compared with the estimated non-thermal flux values from the model. We have employed the publicly available 
AGN-SED-tool (SSC/EC Simulator)\footnote{http://www.isdc.unige.ch/sedtool/PROD/SED.html} to derive the best 
fit parameters for the broadband spectral energy distributions of the blazar 3C 279 during three epoch as 
described below. 

\subsection{Pre-flare (June 1-13, 2015)}
\begin{figure}
\epsscale{0.75}
\plotone{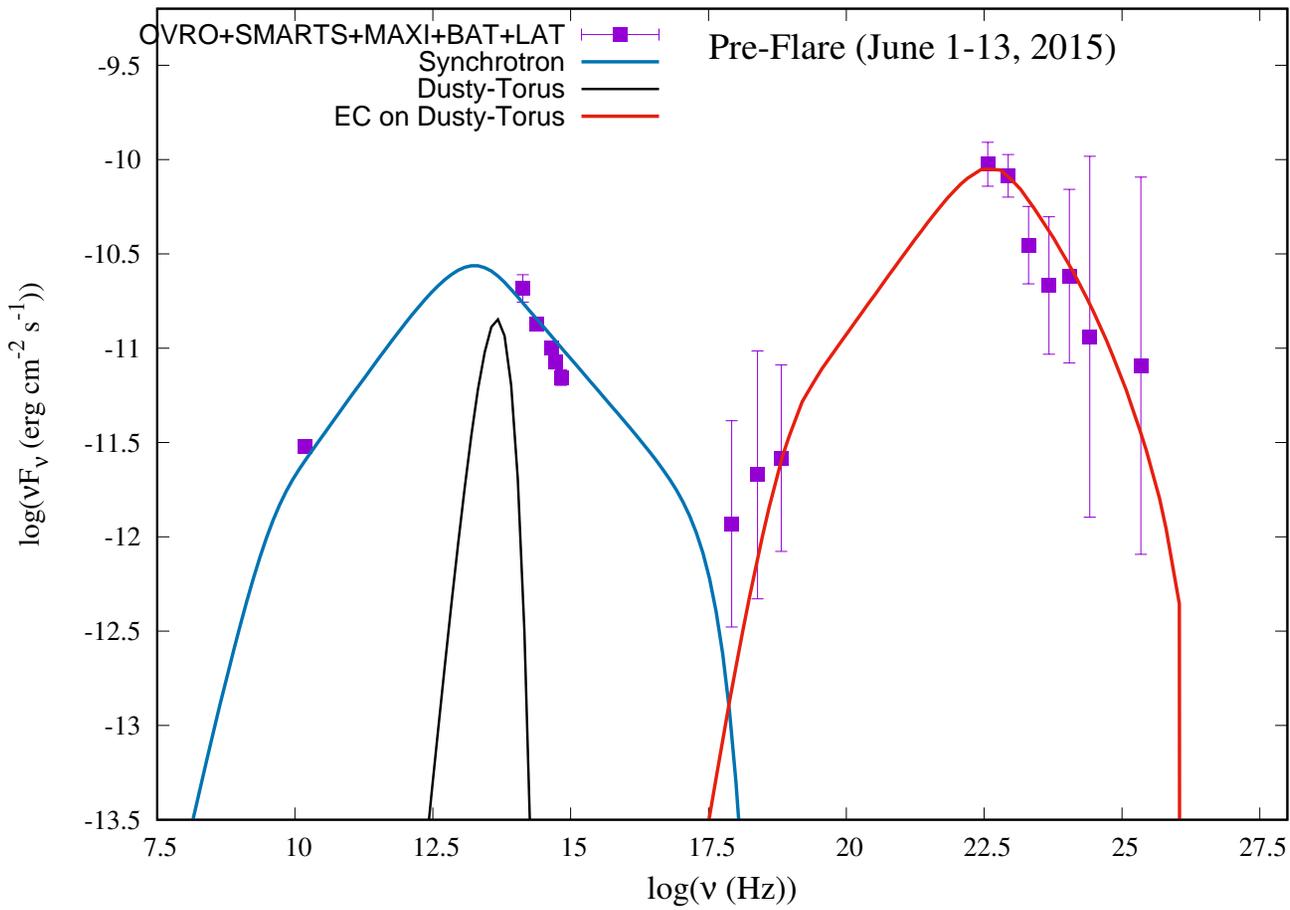}
\caption{Broadband spectral energy distribution of the blazar 3C 279 in the low activity state before the flaring 
	episode. The multi-wavelength flux points at radio, optical, X-ray and high energy $\gamma$-ray energies 
	correspond to the time-averaged near simultaneous observations during June 1-13, 2015 from the OVRO, SMARTS, 
	MAXI \& BAT and LAT respectively.} 
\label{fig2}
\end{figure}

The pre-flare epoch is defined to characterize the low activity state of the source prior to the flaring 
episode. For this epoch,  we have used near simultaneous multi-wavelength observations available during  
June 1-13, 2015 from the OVRO, SMARTS, 	MAXI \& BAT and LAT. The time-averaged flux points obtained from 
these observations are shown in Figure \ref{fig2}. We have employed the one zone leptonic model described above 
to reproduce the radio, optical, X-ray and high energy $\gamma$-ray emissions during the pre-flare epoch. 
The model curves corresponding to the synchrotron and inverse Compton scattering of the external photons from 
the dusty torus (EC on DT) during the low activity state of 3C 279 are depicted in Figure \ref{fig2}. We observe that 
the model satisfactorily reproduces the time-averaged broadband flux measurements in the low activity state of the source. 
The best fit model parameters  obtained from the SED modelling are given in column 3 of the Table \ref{table1}. 

\subsection{Flare (June 16, 2015)}
\begin{figure}
\epsscale{0.75}
\plotone{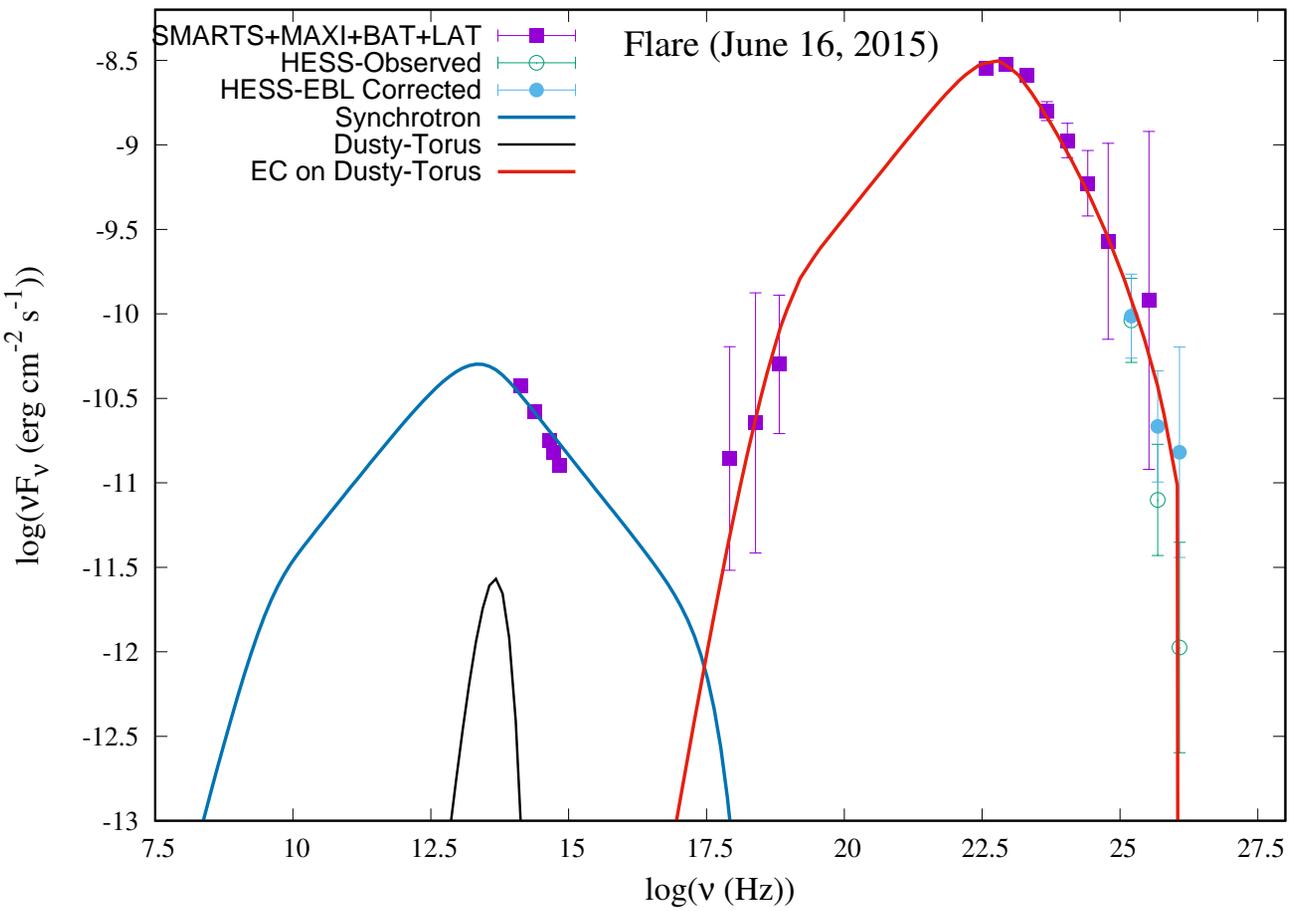}
\caption{Broadband SED of 3C 279 during the giant outburst on June 16, 2015. 
	The broadband flux measurements at optical, X-ray and $\gamma$-ray energies are obtained from 
	the near simultaneous observations on June 16, 2015 from the SMARTS, MAXI \& BAT, LAT and H.E.S.S. 
	respectively. The very high energy $\gamma$-ray flux points from the H.E.S.S. are corrected for 
	the EBL absorption using \citep{Franceschini2008} model.} 
\label{fig3}
\end{figure}

The flare epoch corresponds to the period of a giant outburst of the blazar 3C 279 detected by the \emph{Fermi}-LAT on June 
16, 2015. We have obtained the contemporaneous data available from the SMARTS, MAXI \& BAT and LAT in the optical, X-ray and 
high energy $\gamma$-ray bands respectively. Fortunately, very high energy data from the H.E.S.S. observations are also available 
during the night of the flaring episode. The observed very high energy above 66 GeV is described by a power law of the 
form \citep{Hess2019} 
\begin{equation}
	\frac{dN}{dE}~=~(2.5 \pm 0.5)\times10^{-9} \left(\frac{E}{98 \rm GeV}\right)^{-4.2 \pm 0.3}~~\rm{ph~cm^{-2}~s^{-1}~TeV^{-1}}
\end{equation}
We have derived the very high energy flux points at three energies 66 GeV, 200 GeV, and 500 GeV from the above relation 
using the error propagation method to construct the broadband SED of the source during the outburst. These flux points 
have also been corrected for absorption due to EBL via photon-photon pair production following the methodology described in 
\citep{Singh2014} corresponding to the EBL model proposed by \cite{Franceschini2008}. The flux points are converted to the 
$\nu F_\nu$ (erg~cm$^{-2}$~s$^{-1}$) values for the SED representation using the relation
\begin{equation}
	E^2\frac{dN}{dE}~\equiv~\nu F_\nu
\end{equation}	
The contemporaneous multi-wavelength flux points along with the SED model curves are shown in Figure \ref{fig3}. It is evident 
from the figure that the leptonic one zone model reproduces the broadband emission from the FSRQ 3C 279 during the flaring activity 
very well. The corresponding best fit model parameters are reported in column 4 of the Table \ref{table1}.

\subsection{Post-flare (June 18-30, 2015)}
\begin{figure}
\epsscale{0.75}
\plotone{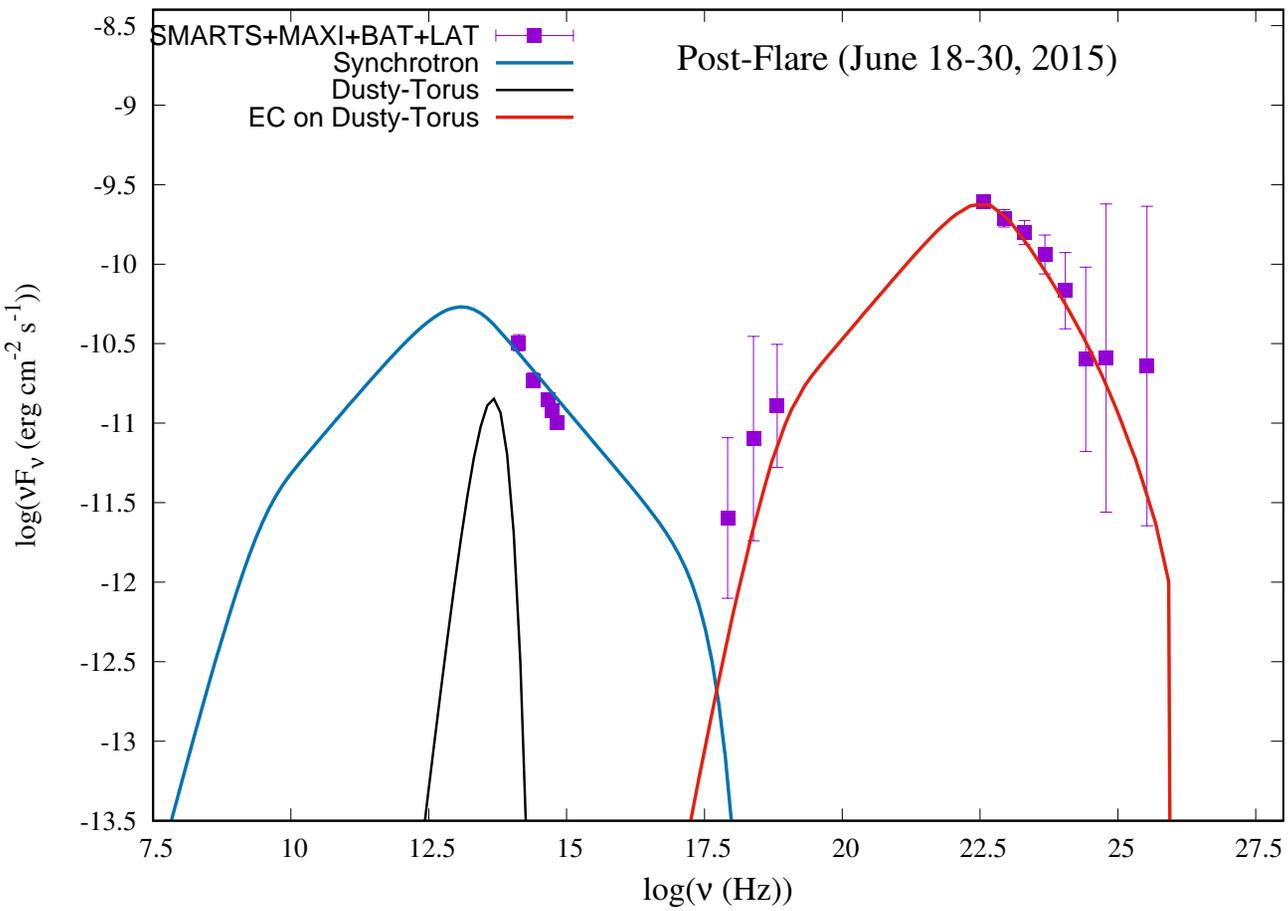}
\caption{Same as Figure \ref{fig1} for the period June 18-30, 2015.} 
\label{fig4}
\end{figure}

The flaring activity of the source decays to low emission state on June 17, 2015. Therefore, we have defined the post-flare epoch
of the blazar 3C 279 during the period June 18-30, 2015 to characterize the low activity state of the source after the vanishing of the 
outburst. The broadband emissions measured in optical, X-ray and high energy $\gamma$-ray bands from the near simultaneous 
observations using SMARTS, MAXI \& BAT and LAT during the post-flare epoch are depicted in Figure \ref{fig4}. The model curves 
in Figure \ref{fig4} for reproducing the broadband emission from 3C 279 are in good agreement with the measured flux 
points. The best fit parameters of the model corresponding to the post-flare epoch are summarized in column 5 of the 
Table \ref{table1}.

\section{Discussion}
Near simultaneous observations of the giant $\gamma$-ray flaring activity of the blazar 3C 279 by the 
\emph{Fermi}-LAT and H.E.S.S. telescopes on June 16, 2015 provide a very important data set for exploring  
the parameter space of the blazar emission model. The minimum variability timescale of $\sim$ 5 minutes 
measured by the \emph{Fermi}-LAT during the outburst of the source characterizes the blazar 3C 279 as 
one of the blazars with the fastest variability in the GeV-TeV energy regime \citep{Ackermann2016,Hess2019}. 
We have used a simple one zone leptonic model to reproduce the non-thermal broadband emission from the 
source during low and high activity states. The sets of model parameters derived from the SED modelling 
(Table \ref{table1}) represent one of the possible combinations for the source activity. A simple 
comparison of the individual model parameters obtained for three different epochs suggests that the 
electron number density ($N$) in the emission region changes significantly during the flaring episode. 
However, the electron energy distribution is described by a broken power law with spectral indices 
$p~\sim$ 2.2 and $q~\sim$ 3.7 at low and high energies respectively. If the broken power law for electron 
energy distribution is simply attributed to the radiative cooling effects, the spectral indices should satisfy 
the condition $q=p+1$. A deviation from this condition indicates a more complex situation where different acceleration 
or cooling processes dominate in different energy bands. In the present study, the values of the spectral indices 
broadly satisfy the condition of radiative cooling break. The energy density of the radiating particles or electrons ($U_e$)
and magnetic field ($U_B$) in the comoving frame can be estimated as 
\begin{equation}
	U_e~=~\int_{\gamma_{min}}^{\gamma_{max}} m_e c^2 \gamma n(\gamma) d\gamma 
\end{equation}	
and 
\begin{equation}
	U_B~=~\frac{B^2}{8\pi}. 
\end{equation}
From Table \ref{table1}, the strength of the magnetic field in the emission region is found to be $\sim$ 0.37 G during all the three epochs, 
whereas the electron number density attains a maximum value of $\sim~ 10^4$ during the flaring activity of the blazar 3C 279. The estimated 
values of $U_e$ and $U_B$ for three epochs of the source activity are listed in Table \ref{table2}. We observe that the particle and 
magnetic energy densities are near the equipartition in the low activity states, whereas during the flaring episode, the two quantities are 
far from the equipartition suggesting that the jet is particle or matter dominated. The flaring activity can be attributed to the sudden 
increase in the electron density in the emission region. Also, the size of the emission region during the outburst is smaller than that 
in the low activity states of the source before and after the flare (Table \ref{table1}). This suggests that the non-thermal emission from 
the inner region of the blob with relatively higher electron energy density dominates the broadband emission measured during the $\gamma$-ray 
flaring activity of 3C 379 on June 16, 2015. The strong flaring activity due to the sudden increase in the particle energy density has also 
been modelled in individual blazars like Mrk 421 using a time dependent leptonic model \citep{Singh2017}. The equipartition condition between 
the radiating particle and magnetic field energy densities offers a minimum power solution to the blazar jet emission. It becomes highly 
uncertain when magnetic field energy density is coupled with the total baryonic energy density of the jet.
\par
The synchrotron peak frequency $\nu_p~\sim$ 3.2$\times$10$^{13}$ Hz obtained from the SED modelling during the three epochs supports 
the classification of 3C 279 as a low-synchrotron-peaked blazar. The observed synchrotron peak frequency can be expressed in terms 
of the model parameters as 
\begin{equation}
	\nu_p~=~\left(\frac{\delta}{1+z}\right)\frac{\gamma_b^2 e B}{2\pi m_e c}
\end{equation}	
where $e$ is the electronic charge. The $\gamma$-ray emission from the source can be attributed to the inverse Compton scattering of the 
external photons originating from the dusty torus with a temperature of 870 K located at a distance of $\sim$ 6$\times$10$^{19}$ cm from 
the central engine. This dusty torus reprocesses a fraction ($\sim$ 20$\%$) of the accretion disc luminosity in the IR regime. 
The Comptonization of IR seed photons emitted from the dusty torus takes place in the Thomson regime. The non-thermal emissions from 
blazars demand very large energies and extreme powers associated with the relativistic jets. The physics of powering the jets in 
astrophysical systems like blazars hosting a SMBH at the center is not clearly understood. Based on the various parameters derived from 
the SED modelling of the blazar 3C 279, the jet kinetic power ($P_j$) in the stationary frame of the host galaxy of the source can be 
obtained from the relation \citep{Celotti2008}
\begin{equation}
	P_j~=~\pi R^2 \Gamma^2 c (U_e + U_B + U_p)
\end{equation}
where $U_p$ is the comoving energy density of the cold protons in the jet plasma which can be calculated as 
\begin{equation}
	U_p~=~m_p c^2 \int n(\gamma) d\gamma , 
\end{equation}	
$m_p$ being the rest mass of the proton. The estimated values of the jet power and other associated quantities from 
the best fit model parameters are given in Table \ref{table2}. It is observed that the jet power and radiative power 
($P_r$) corresponding to the non-thermal emission are highest during the flaring epoch. The radiative power is found to 
be less than the jet power, indicating that only a small fraction of the jet kinetic power is dissipated in the form of 
observed radiation. This is consistent with the general characteristics of the blazar jet emission. The Eddington luminosity 
for a SMBH of mass $M$ is given by 
\begin{equation}
	L_{Edd}~=~1.25 \times 10^{38} \left(\frac{M}{M_\odot}\right)~~{\rm erg~s^{-1}}
\end{equation}	
where $M_\odot$ is the solar mass. For the blazar 3C 279, $M$ is estimated to be in the range of (3-8)$\times$10$^8$ M$
_\odot$ \citep{Nilsson2009}. Therefore, the maximum Eddington luminosity for 3C 279 is $\sim$ $10^{47}$ erg~s$^{-1}$, which is 
larger than the jet power reported in Table \ref{table2}. This indicates that the non-thermal emission from 3C 279 takes place at 
sub-Eddington limit. The model parameter space obtained in the present study is broadly consistent with the previous values 
in the literature \citep{Sunder2018,Hayashida2015,Sunder2012}. Recently, \cite{Shah2019} have studied the flaring episode of 
3C 279 observed in January 2018 and report that the increase in the value of $\Gamma$ causes flux enhancement during the outburst. 
\cite{Bottcher2013} have shown that the hadronic models with a very large power in relativistic protons, high magnetic field strength 
and extreme jet power cannot satisfactorily fit the broadband SED of 3C 279 due to the spectral break at GeV energies in the 
observed spectrum. A clumpy jet model with strings of compact plasmoids is also proposed to interpret the bright and fast $\gamma$-ray 
flares of the blazar 3C 279 \citep{Vittorini2017}. In this model, low energy emission in optical-UV bands is attributed to the synchrotron 
radiation due to relativistic electrons accelerated around the plasmoids. The synchrotron radiation is partially reflected back by a 
string which acts like a moving mirror for the approaching companions. Resulting which, shrinking gap transient overdensities of seed 
photons are build up around the plasmoid and Comptonization of these seed photons produces high energy $\gamma$-rays during the 
flaring episodes \citep{Vittorini2017}. 

\section{Conclusion}
In this paper, we study the broadband spectral energy distribution of 3C 279 using the multi-wavelength observations 
during June 1-30, 2015. We have produced the SEDs of the source for three epochs corresponding to the low and high 
activity states under the framework of a simple one zone leptonic model. We find that the non-thermal broadband emission 
from the jet of blazar 3C 279 during the whole period can be reproduced by an electron energy distribution described by a broken 
power law. During the low activity states of the source before and after the giant outburst on June 16, 2015, the magnetic and 
electron energy densities approximately satisfy the equipartition condition. The giant flaring activity in the source is caused by 
a sudden increase in the electron energy density in the inner region of the emitting blob. The size of the emission region during the 
flare is smaller than that in the low activity state by a factor of $\sim$ 4. The enhancement in the electron energy density can be 
attributed to the transient injection of relativistic electrons in the inner region of the jet. At low energies, the radiation is 
produced by the synchrotron process, and inverse Compton scattering of the seed photons from the dusty torus dominates at X-ray and 
$\gamma$-ray energies. The high energy emission from 3C 279 due to the Comptonization of the IR photons from dusty torus indirectly 
confirms the presence of hot dust in blazars.
\acknowledgments
We thank the anonymous reviewer for his/her suggestions and comments to improve the contents of this study.
This research has made use of data from the OVRO 40-m monitoring program 
(Richards, J. L. et al. 2011, ApJS, 194, 29) which is supported in part by NASA grants 
NNX08AW31G, NNX11A043G, and NNX14AQ89G and NSF grants AST-0808050 and AST-1109911.
This paper has made use of up-to-date SMARTS optical/near-infrared light curves that 
are available at www.astro.yale.edu/smarts/glast/home.php.
This research has made use of MAXI data provided by RIKEN, JAXA and the MAXI team.
We acknowledge the use of Swift/BAT transient monitor results provided by the Swift/BAT team 
in this research.


\begin{table}
\begin{center}
\caption{Summary of the best fit parameters for the broadband spectral energy distributions of the FSRQ 3C 279 
	during three epochs from the simple one zone leptonic emission model for blazars.}
\vspace{1.0cm}
\begin{tabular}{lcllll}
\tableline\tableline
Parameters (Units) 		&Symbol  	&Pre-Flare		&Flare 		&Post-Flare\\
\tableline
Redshift                        &z              &0.536			&0.536			&0.536\\
Size of emission region (cm) 	&R 		&9.3$\times$10$^{16}$ 	&2.1$\times$10$^{16}$	&9.3$\times$10$^{16}$\\
Jet viewing angle ($^\circ$)    &$\theta$       &2			&2			&2\\
Bulk Lorentz factor		&$\Gamma$  	&21			&21			&21\\
Magnetic Field (G)		&B		&0.37			&0.37			&0.37\\
Low energy spectral index       &p		&2.18			&2.18			&2.18\\
High energy spectral index      &q  		&3.68			&3.82			&3.82\\
Minimum electron Lorentz factor &$\gamma_{min}$ &10			&10			&10\\	
Maximum electron Lorentz factor &$\gamma_{max}$ &10$^5$			&10$^5$			&10$^5$\\	
Electron Lorentz factor at break &$\gamma_b$	&800			&950			&750\\
Electron number density (cm$^{-3}$) &N  	&70   			&10$^4$   		&200\\
Accretion disc luminosity (erg~s$^{-1}$)&L$_{disc}$ &1.9$\times$10$^{46}$&1.9$\times$10$^{46}$	&1.9$\times$10$^{46}$\\
Accretion disc temperature (K) &T$_{disc}$	&10$^5$			&10$^5$			&10$^5$\\
Location of dusty torus (cm)     &R$_{DT}$	&6.1$\times$10$^{19}$	&6.1$\times$10$^{19}$   &6.1$\times$10$^{19}$\\   
Dust temperature (K)		 &T$_D$		&870			&870			&870\\
\tableline
\end{tabular}
\label{table1}
\end{center}
\end{table}
\begin{table}
\begin{center}
\caption{Few important source parameters derived from the spectral energy distribution modelling of the FSRQ 3C 279 
	 during three epochs.}
\vspace{1.0cm}
\begin{tabular}{lcllll}
\tableline\tableline
Parameters (Units) 		       &Symbol  	&Pre-Flare		&Flare 		&Post-Flare\\
\tableline
Doppler factor			       &$\delta$  	&27			&27			&27\\
Electron energy density (erg~cm$^{-3}$) &U$_e$  	&2.2$\times$10$^{-3}$  &3.2$\times$10$^{-1}$   &6.3$\times$10$^{-3}$\\
Proton energy density   (erg~cm$^{-3}$) &U$_p$          &1.1$\times$10$^{-2}$  &1.5		       &3.0$\times$10$^{-2}$\\
Magnetic energy density  (erg~cm$^{-3}$) &U$_B$ 	&5.4$\times$10$^{-3}$  &5.4$\times$10$^{-3}$   &5.4$\times$10$^{-3}$\\
Synchrotron peak frequency (Hz)        &$\nu_p$ 	&3.2$\times$10$^{13}$  &3.2$\times$10$^{13}$   &3.2$\times$10$^{13}$\\
Jet power	(erg~s$^{-1}$)           &P$_j$		&6.5$\times$10$^{45}$  &3.5$\times$10$^{46}$   &1.4$\times$10$^{46}$\\
Radiative power (erg~s$^{-1}$)		 &P$_r$		&1.7$\times$10$^{44}$  &4.4$\times$10$^{45}$   &4.2$\times$10$^{44}$\\
Total power     (erg~s$^{-1}$)		 &P		&6.7$\times$10$^{45}$  &3.8$\times$10$^{46}$   &1.5$\times$10$^{46}$ \\
\tableline
\end{tabular}
\label{table2}
\end{center}
\end{table}
\end{document}